\documentclass[twocolumn,twoside,slac_two]{revtex4}
\usepackage{graphicx}
\usepackage{mathrsfs}
\usepackage{fancyhdr}
\usepackage{hyperref}
\usepackage{xspace}
\hypersetup{
    colorlinks=true,
    urlcolor=magenta,
}
\def\EeV{\ifmmode {\mathrm{Ee\kern -0.07em V}}\else
                   \textrm{Ee\kern -0.07em V}\fi}
\def\TeV{\ifmmode {\mathrm{Te\kern -0.07em V}}\else
                   \textrm{Te\kern -0.07em V}\fi}
\def\eV{\ifmmode {\mathrm{\ e\kern -0.07em V}}\else
                   \textrm{e\kern -0.07em V}\fi}
\def\gcm{\ensuremath{\mathrm{g/cm}^2}\xspace}
\def\xmax{\ensuremath{X_\mathrm{max}}\xspace}
\def\sigmaXmax{\ensuremath{\sigma(X_\mathrm{max})}\xspace}
\def\meanXmax{\ensuremath{\langle X_\mathrm{max}\rangle}\xspace}
\newcommand{\energy}[1]{\ensuremath{10^{#1}}\,\eV}
\def\Sibyll{\textsc{Sibyll2.1}\xspace}
\def\Epos{\textsc{Epos-LHC}\xspace}

\def\QgIINew{\textsc{QGSJetII-04}\xspace}

\begin{document}

\title{Measurements of the depth of maximum of air-shower profiles at the Pierre Auger Observatory and their composition implications}

\author{Vitor de Souza$^1$ for the Pierre Auger Collaboration$^2$}
\affiliation{1 - Instituto de F\'isica de S\~ao Carlos, Universidade de S\~ao Paulo, S\~ao Carlos, Brasil}
\affiliation{2 - Full author list: http://www.auger.org/archive/authors\_2016\_09.html.}

\begin{abstract}
Air-showers measured by the Pierre Auger Observatory were analyzed in order to extract the depth of maximum (\xmax).The results allow the analysis of the \xmax distributions as a function of energy ($> 10^{17.8} \eV$). The \xmax distributions, their mean and standard deviation are analyzed with the help of shower simulations with the aim of interpreting the mass composition. The mean and standard deviation were used to derive $<ln A>$ and its variance as a function of energy. The fraction of four components (p, He, N and Fe) were fit to the \xmax distributions. Regardless of the hadronic model used the data is better described by a mix of light, intermediate and heavy primaries. Also, independent of the hadronic models, a decrease of the proton flux with energy is observed. No significant contribution of iron nuclei is derived in the entire energy range studied.
\end{abstract}

\maketitle

\thispagestyle{fancy}

\section{Introduction}

Major advances in the study of the origin of cosmic rays depend on the determination of the abundances of primaries. The composition of the ultra-high cosmic ray population reaching the Earth is a key feature in the puzzle of production and propagation of these particles.

This contribution presents the composition scenario built by the Pierre Auger Observatory~\cite{Aab:2014kda,Aab:2014aea,icrc2015,uhecr2014} based on \xmax measurements. The use of fluorescence telescopes allows the most accurate determination of \xmax in each shower. The development of a robust data analysis chain leads to a minimum biased selection of events and to a high resolution determination of \xmax. The traditional analysis of the moments of the \xmax distribution as a function of energy confirms previous results~\cite{bib:auger:xmax:prl}. Steps forward were also taken: the full distribution is interpreted with the use of shower simulations and an extension to lower energies was achieved with the use of HEAT~\cite{heat}.

This paper is based in part on recent Auger publications~\cite{Aab:2014kda,Aab:2014aea,icrc2015,uhecr2014} in which shower depth data were analyzed and interpreted. A summary of the salient points and conclusions is presented here, and the reader is referred to the publications for full details. Section~\ref{sec:data} describes the data samples used. Section~\ref{sec:results} presents the results and section~\ref{sec:conclusion} draws the conclusions.

\begin{figure*}[!t]
  \centering
    \includegraphics[width=0.8\textwidth]{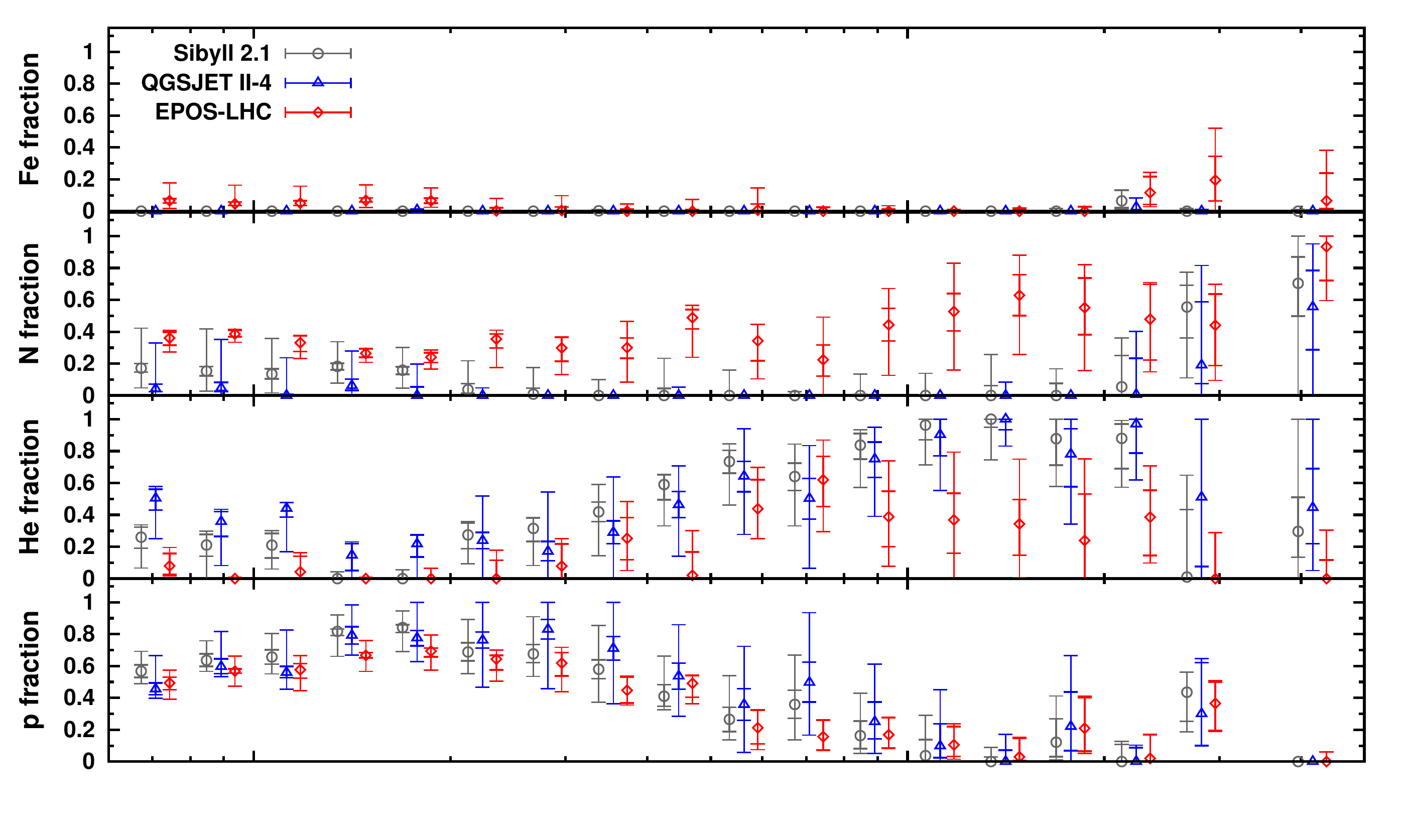}

  \vspace*{-10mm}
  \includegraphics[width=0.8\textwidth]{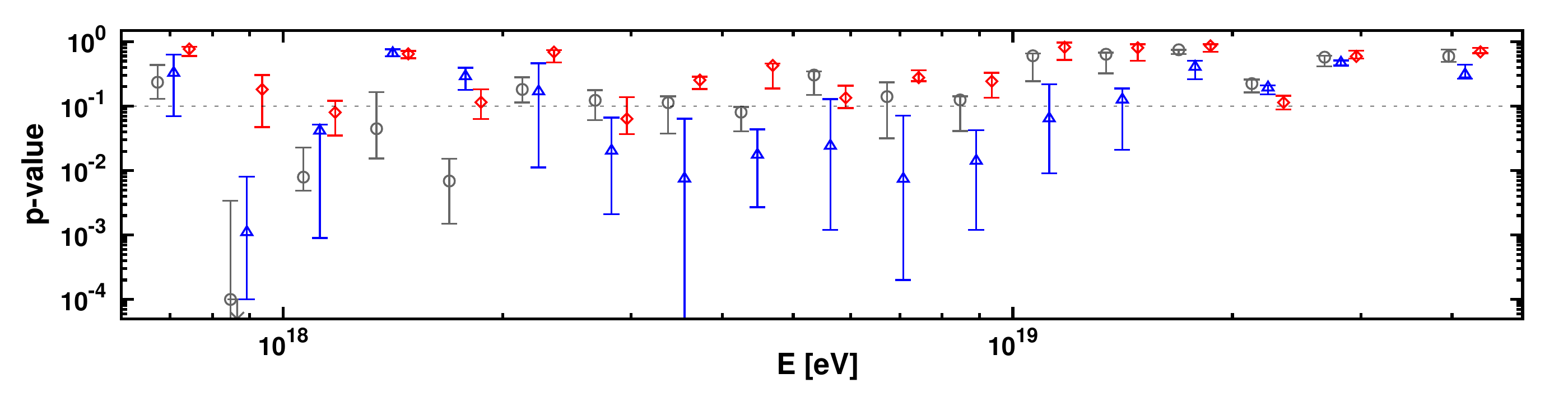}

  \caption{Fraction of each fitted primary species as a function of energy. Three hadronic interaction models are shown.}
  \label{fig:fraction}
\end{figure*}

\section{Data Samples}
\label{sec:data}

The data analyzed here consist of shower candidates measured by the Pierre Auger Observatory from December 1st, 2004 to December 31st, 2012 with the FD telescopes and also events recorded with the HEAT telescopes from June 1st 2010 to August, 15th 2012. A sequence of selection criteria are applied to this data set in order to remove poor quality events. First, good data-taking conditions are required, as determined from a record of information on telescope performance, and stored during operation. In this step, the atmospheric conditions are also qualified and bad periods are removed from further analysis. After the basic operational quality selection, showers are reconstructed according to the procedure described in reference~\cite{Aab:2014kda}. The longitudinal profile is fitted, thus allowing energy and \xmax to be determined. Quality cuts are applied to guarantee the success of the reconstruction procedure. A fiducial selection is applied in order to have constant acceptance along the full \xmax range, thus minimizing the need to correct the final results.

\section{Results}
\label{sec:results}

\subsection{\xmax distribution and its fit}

The data were divided into eighteen energy interval starting at \energy{17.8}. These distributions were fitted using Monte Carlo simulation predictions. \xmax MC templates were generated using CONEX v4r37~\cite{bib:conex:1,bib:conex:2}. Three hadronic interaction models were used: \Epos~\cite{bib:epos}, \QgIINew~\cite{bib:qgsjet} and \Sibyll~\cite{bib:sibyll}. For each species (p, He, N and Fe) $2 \times 10^4$ showers were simulated for each energy bin. The \xmax distribution for each specie and energy bin is corrected by acceptance.

The measured \xmax distribution is fitted using a binned maximum-likelihood method to search for the best combination of species which matches the data. The \xmax distributions were fitted with two (proton and iron nuclei), three (proton, nitrogen and iron nuclei) and four (proton, helium, nitrogen and iron nuclei) primaries species, respectively. The combination of two primaries (p+Fe) resulted in a poor description of the data. The fits with three and four primaries resulted in  good descriptions of the data.

The corresponding abundances of the four primary species (p+He+N+Fe) fit of the \xmax distribution are shown in figure~\ref{fig:fraction}. The percentage of each primary is shown as a function of energy for all hadronic interaction models considered in this analysis. Similar plots for two and three primary species can be found in reference~\cite{Aab:2014aea}.

\begin{figure}[!]
    \centering
    \includegraphics[width=0.5\textwidth]{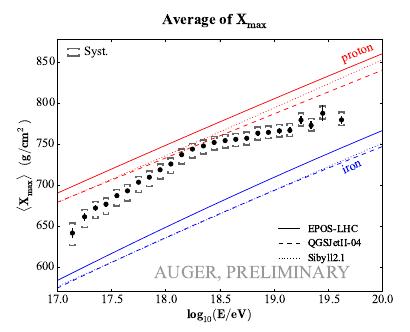}
    \includegraphics[width=0.5\textwidth]{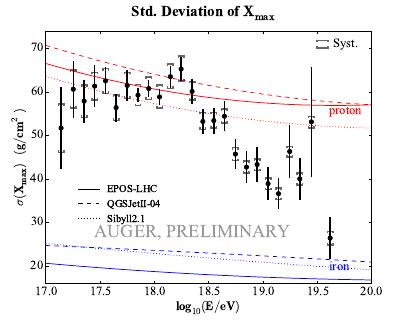}
    \caption{Evolution of the \xmax moments with energy. Black markers are the data. Lines show the predictions of hadronic interaction models.}
    \label{fig:moments}
\end{figure}

\begin{figure*}[!]
  \centering
  \includegraphics[width=\textwidth]{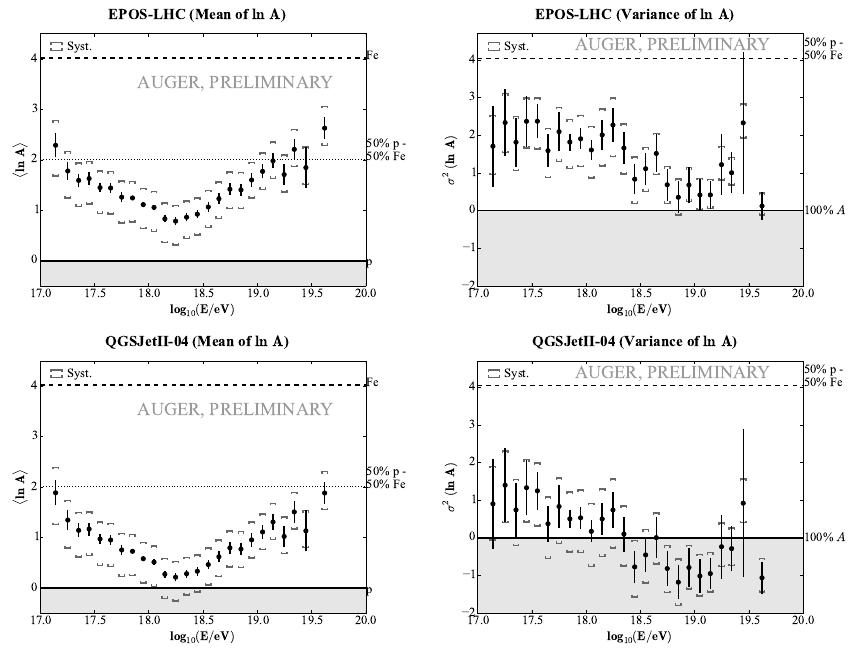}
  \caption{\emph{Upper panel}: Mean logarithmic mass number ($ln (A)$) as a function of energy. \emph{Lower panel}: Variance of $ln (A)$. Each column shows the calculation based on one hadronic interaction model.}
  \label{fig:lna}
\end{figure*}

\subsection{\xmax Moments and $ln (A) $}

The calculation of the first and second moments of the \xmax distribution is a traditional way to study the evolution of the composition with energy. The moments of the distributions are shown in figure~\ref{fig:moments}. The calculation of the moments was done using three techniques as explained in reference~\cite{Aab:2014kda}. The techniques used to calculate the moment all agree within the uncertainties, which lends credence to the results and conclusions.

The data shown in figure~\ref{fig:moments} are independent of Monte Carlo models. However their interpretation in terms of composition depends on hadronic interaction models. The moments predicted by the most used hadronic interaction models are also shown in figure~\ref{fig:moments} for comparison.

The \meanXmax and \sigmaXmax can be converted to the first two moments of the distribution of the logarithm of the primary particle mass number~\cite{bib:linsley}. The result of the conversion is shown in figure~\ref{fig:lna}. Each column of the figure shows the result of the conversion for one hadronic interaction model.

\section{Conclusions}
\label{sec:conclusion}

The measurement of the \xmax distribution for energies between \energy{17.0} and \energy{19.2} has reached unprecedented resolution. Each energy bin ( $\Delta lg E = 0.1$ ) has more than 100 events, which minimizes the statistical fluctuations to a very low level. The effect of the large number of events can be easily visualized by noting the size of the statistical error bars in the plots of the moments in figure~\ref{fig:moments}. This achievement is credited to the continuous operation of the Pierre Auger Observatory in the last decade.

At the same time, the understanding of the detectors, including the atmosphere, has improved significantly. Small optical effects of the telescopes~\cite{bib:halo}, fluorescence light emission mechanism~\cite{bib:yield} and detector calibration~\cite{bib:time:shift} were scrutinized in order to reduce the systematic uncertainties. This effort resulted in an overall systematic uncertainty in \xmax smaller than 10 \gcm which is the size of the bin used in the \xmax distributions.

Little room for improvement is left in this energy range (from \energy{17.0} to \energy{19.2}) based only on \xmax measurements. The continuous acquisition of data and further reduction of the systematics are not going to bring new information due to the limitations imposed by intrinsic shower fluctuations and by the characterization of the atmosphere. Nevertheless, this is a very important energy range. The transition from galactic to extragalactic predominance on the flux might happen in this energy interval and it is expected that a composition evolution should reveal it. An upgrade of the Pierre Auger Observatory is proposed to measure the muon content of the showers in order to improve the quality of the composition determination by adding extra information to \xmax~\cite{prime}.

The \xmax data show only one clear feature at $E=10^{18.27} \eV$. This feature is seen in the \meanXmax and in the \sigmaXmax evolution with energy. Whether the feature at $E=10^{18.27} \eV$ is caused by astrophysical phenomena or by new particle physics is yet to be understood. The interpretation of the \xmax data relies heavily on the hadronic interaction models and two interpretations are shown in this paper: $ln (A)$ (figure~\ref{fig:lna}) and the abundance fit (figure ~\ref{fig:fraction}).

The interpretation based on the abundance fit makes use of the full \xmax distribution. It uses the full power of the data because every single \xmax measurement contributes to the interpretation. However, for the same reason it depends heavily on the hadronic models. The \xmax distribution is mainly shaped by cross-section, elasticity and multiplicity of the first interactions of the shower. Small changes in these parameters contribute to major distortions of the \xmax distribution truncating its tail and changing the inclination of the fall~\cite{bib:ulrich}.

The interpretation based on  $ln (A)$ and its variance is a conversion of the moments of the distribution based on shower simulations. It makes less use of the full data than the fit of the abundances. The interpretation based on $ln (A)$ makes use only of the first two moments of the \xmax distribution instead of using the full distribution. However, it is reasonable to accept that the hadronic interaction models are able to describe the moments of the distribution more accurately than the full distribution can. Therefore, this interpretation is less sensitive to the details of the models.

Figures~\ref{fig:fraction} and~\ref{fig:lna} summarize our interpretation of the data. These two plots present probably the best information about composition available for the near future concerning cosmic rays with energy between \energy{17.0} and \energy{19.2} measured in the Southern hemisphere. Both interpretations agree that the composition starts intermediate at \energy{17}, gets very light at \energy{17.8} and reaches a minimum at \energy{18.27} and becomes heavier again. Up to \energy{18.6}: a) the average $ln (A)$ is below one ($ln (A) = 1.38$ for He), b) the proton fraction is above 50\% and c) the fraction of protons and He taken together is above 90\%. A rather constant percentage of intermediate mass nuclei labeled as Nitrogen is needed to describe the data and a small increase of the Nitrogen fraction is suggested by the abundance analysis for energies above \energy{19}. Almost no iron nuclei are needed in the entire energy range ($E > 10^{17.8} \eV$). Therefore the feature seen at $E=10^{18.27} \eV$ is caused by the decrease of the proton flux and not by the increase of the flux of heavy primaries.

\bigskip 
\begin{acknowledgments}
The author acknowledges financial support from FAPESP (2010/07359-6,2014/19946-4) and CNPq (Brazil).
\end{acknowledgments}

\bigskip 

\begin{thebibliography}{99} 

\bibitem{Aab:2014kda}
  A.~Aab {\it et al.}  [Pierre Auger Collaboration],
  Phys.\ Rev.\ D {\bf 90}, no. 12, 122005 (2014).


\bibitem{Aab:2014aea}
  A.~Aab {\it et al.}  [Pierre Auger Collaboration],
  Phys.\ Rev.\ D {\bf 90}, no. 12, 122006 (2014).

\bibitem{icrc2015}
  The Pierre Auger Collaboration, arXiv 1509.03732.

\bibitem{uhecr2014}
The Pierre Auger Collaboration, JPS Conf. Proc.(UHECR 2014) 9, 010015 (2016).

\bibitem{heat}
T. Hermann-Josef Mathes, for the Pierre Auger Collaboration, Proc. 32nd ICRC, Beijing, China 3, 149 (2011); arXiv:1107.4807.

\bibitem{bib:auger:xmax:prl}
  J.~Abraham {\it et al.}  [Pierre Auger Collaboration],
  Phys.\ Rev.\ Lett.\  {\bf 104}, 091101 (2010).

\bibitem{bib:conex:1}
  T.~Bergmann {\it et al.},
  Astropart.\ Phys.\  {\bf 26}, 420 (2007) [astro-ph/0606564].

\bibitem{bib:conex:2}
  T.~Pierog {\it et al.},
  Nucl.\ Phys.\ Proc.\ Suppl.\  {\bf 151}, 159 (2006)
  [astro-ph/0411260].

\bibitem{bib:qgsjet}
  S.~Ostapchenko,
  Phys.\ Rev.\ D {\bf 83}, 014018 (2011).

\bibitem{bib:sibyll}
  E.~J.~Ahn {\it et al.},
  Phys.\ Rev.\ D {\bf 80}, 094003 (2009).

\bibitem{bib:epos}
  T.~Pierog {\it et al.},
  arXiv:1306.0121 [hep-ph].

\bibitem{bib:linsley}
  J.~Linsley, Proc. 15th ICRC 12 (1977) 89, Plovdiv, Bulgaria.

\bibitem{bib:halo}
J.~B\"auml {\it et al.} [Pierre Auger Collaboration], Proc. 33rd ICRC (2013), arXiv:1307.5059, Rio de Janeiro, Brazil.

\bibitem{bib:yield}
M.~Ave {\it et al.} [AIRFLY Collaboration], Astropart. Phys. 28 (2007) 41.
M.~Ave {\it et al.} [AIRFLY Collaboration], Astropart. Phys. 42 (2013) 90.

\bibitem{bib:time:shift}
P.~Alisson {\it et al.} [Pierre Auger Collaboration], Proc. 29th ICRC 8 (2005) 307, Pune, India.

\bibitem{bib:ulrich}
R.~Ulrich, R.~Engels and M.~Unger, Phys.\ Rev.\ D 83, 054026 (2011).

\bibitem{bib:spec:icrc2013}
A. Schulz {\it et al.} [Pierre Auger Collaboration], Proc. 33rd ICRC (2013), arXiv:1307.5059, Rio de Janeiro, Brazil.

\bibitem{prime}
 The Pierre Auger Collaboration, arXiv:1604.03637.

\end{thebibliography}

\end{document}